\documentclass[runningheads]{llncs}
\usepackage{epsfig}
\usepackage{graphicx}
\usepackage{subfigure}
\usepackage{amsmath}
\usepackage{amssymb}

\usepackage{amsthm}

\usepackage{amsfonts}
\usepackage{mathrsfs}
\usepackage{booktabs}
\usepackage{multirow, eucal}
\usepackage{color}
\usepackage{cite}
\usepackage{wrapfig,lipsum,booktabs}
\usepackage{caption}
\usepackage[vlined,ruled,linesnumbered]{algorithm2e}

\def\red{\textcolor{red}}

\begin{document}
\title{Attention Guided Network for Retinal Image Segmentation\thanks{ This work was done when S. Zhang is intern at CVTE Research. M. Tan (mingkuitan@scut.edu.cn) and Y. Xu (ywxu@ieee.org) are the corresponding authors.}}

\author{
Shihao Zhang$^{1}$ \and
Huazhu Fu$^{2}$ \and
Yuguang Yan$^{1}$ \and
Yubing Zhang$^{4}$ \and
Qingyao Wu$^{1}$ \and
Ming Yang$^{4}$ \and
Mingkui Tan$^{1,3*}$ \and
Yanwu Xu$^{5*}$
}

\authorrunning{S. Zhang et al.}

\institute{South China University of Technology, Guangzhou, China \and
Inception Institute of Artificial Intelligence, Abu Dhabi, UAE \and
Peng Cheng Laboratory, Shenzhen, China \and
CVTE Research, Guangzhou, China\and
Cixi Institute of Biomedical Engineering, Ningbo Institute of Materials Technology and Engineering, Chinese Academy of Sciences, Ningbo, China
~\\
Project page: \red{https://github.com/HzFu/AGNet}
}

\maketitle
\thispagestyle{empty}
\begin{abstract}
Learning structural information is critical for producing an ideal result in retinal image segmentation. Recently, convolutional neural networks have shown a powerful ability to extract effective representations. However, convolutional and pooling operations filter out some useful structural information. In this paper, we propose an Attention Guided Network (AG-Net) to preserve the structural information and guide the expanding operation. In our AG-Net, the guided filter is exploited as a structure sensitive expanding path to transfer structural information from previous feature maps, and an attention block is introduced to exclude the noise and reduce the negative influence of background further. The extensive experiments on two retinal image segmentation tasks (i.e., blood vessel segmentation, optic disc and cup segmentation) demonstrate the effectiveness of our proposed method.
\end{abstract}

\section{Introduction} \label{Sec:Introduction}

Retinal image segmentation plays an important role in automatic disease diagnosis. Compared to general natural images, retinal images contain more contextual structures, e.g., retinal vessel, optic disc and cup, which often provide important clinical information for diagnosis. As the main indicators for eye disease diagnosis, the segmentation accuracy of these information is important.
Recently, convolutional neural networks (CNNs) have shown the strong ability in retinal image segmentation with remarkable performances~\cite{deepvessel,Gu2019,fu2018joint,yan2017skeletal}. Existing CNN based models learn increasingly abstract representations by cascade convolutions and pooling operations. However, these operations may neglect some useful structural information such as edge structures, which are important for retinal image analysis. To address this issue, one possible solution is to add extra expanding paths to merge features skipped from the corresponding resolution levels. For example, FCN~\cite{long2015fully} sums up the upsampled feature maps and the feature maps skipped from the contractive path. And U-Net~\cite{ronneberger2015u} concatenates them and add convolutions and non-linearities. However, these works can not effectively leverage these structural information, which may hamper the segmentation performance. Therefore, it is desirable to design a better expanding path to preserve structural information.

To address this, we introduce guided filter~\cite{he2013guided} as a special expanding path to transfer structural information extracted from low-level feature maps to high-level ones. Guided filter~\cite{he2013guided} is an edge-preserving image filter, and has been demonstrated to be effective for transferring structural information. Different from existing works which use the guided filter at the image level, we incorporate the guided filter into CNNs to learn better features for segmentation.
We further design an attention mechanism in guided filter, called attention guided filter, to remove the noisy components, which are introduced from the complex background by original guided filter. Finally, we propose \textbf{Attention Guided Network (AG-Net)} to preserve the structural information and guide the expanding operation. The experiments on vessel segmentation and optic disc/cup segmentation  demonstrate the effectiveness of our proposed method.

\section{Methodology} \label{Sec:Methodology}

\begin{figure*}[!t]
	\centering
	\includegraphics[width=1\linewidth]{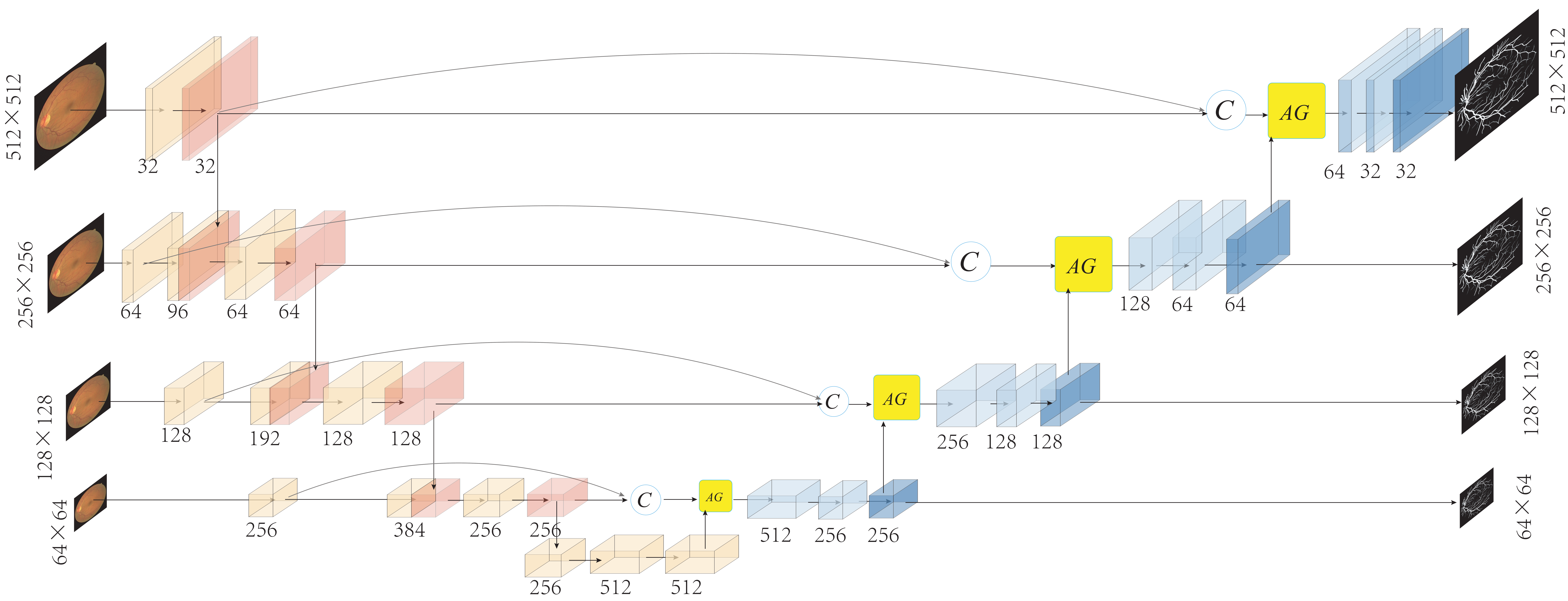}
	\caption{Architecture of proposed AG-Net. Our AG-Net is based on M-Net~\cite{fu2018joint}, which is a multi-scale multi-label segmentation network. The block \textit{AG} represents our attention guided filter and the operator \textit{C} is the concatenation. In our AG-Net, the attention guided filter is used as a structural sensitive skip-connection to replace the skip-connection and upsampling layer for better information fusion. }
	\label{figall}
\end{figure*}

Fig.~\ref{figall} shows the architecture of proposed AG-Net, where M-Net~\cite{fu2018joint} is utilized as the backbone to learn hierarchical representations. We propose attention guided filter into the network, which contains the guided filter and attention block to filter out the noise from the background and address the boundary blur problem caused by upsampling.  The details of our AG-Net are illustrated as follows.

\subsection{Attention Guided Filter}

\begin{figure}[!t]
	\centering
	\includegraphics[width=1\linewidth]{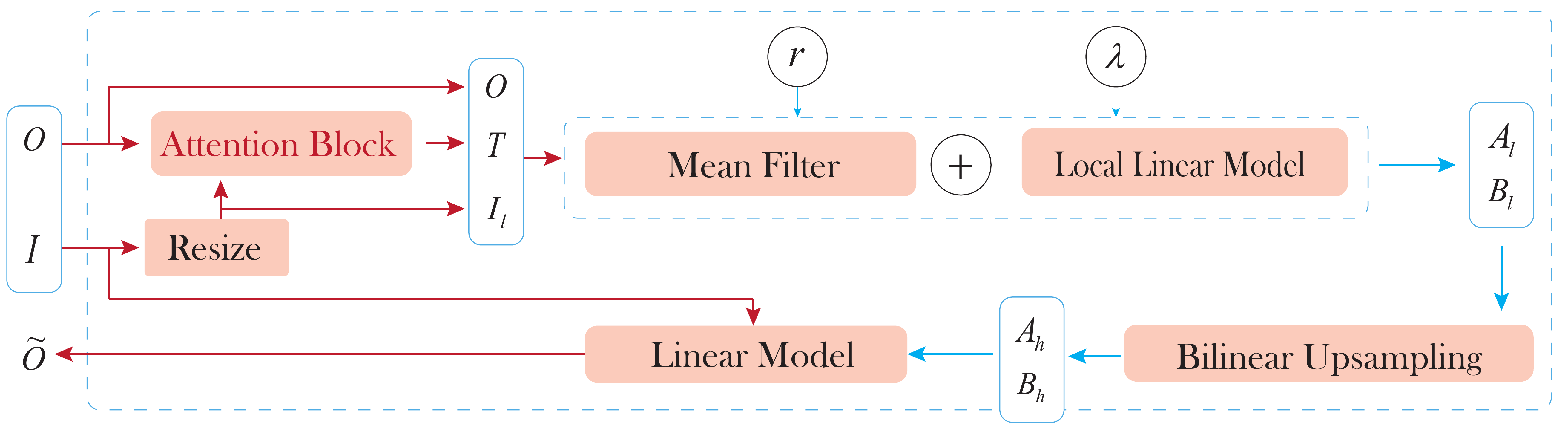}
	\caption{Illustration of the attention guided filter. The attention guided filter first produces the attention map $T$ through the attention block, then calculates $A_l,B_l$ with the attention map $T$, resized guidance feature map ($I_l$), filtering feature map ($O$) and hyperparameter $r, \epsilon$. By using bilinear upsampling $A_l$ and $B_l$, we obtain $A_h$ and $B_h$ for producing the final output $\tilde{O}$ with $I$.}
	\label{figure1}
\end{figure}

The attention guided filter recovers spatial information and merges structural information from the various resolution levels by filtering the low-resolution feature maps with high-resolution feature maps. The inputs include a guidance feature map ($I$), and a filtering feature map ($O$). The output is a high-resolution feature map $\tilde{O}$. The attention feature map $T$ is produced by an attention block.
As shown in Fig.~\ref{figure1}, the attention guided filter firstly downsamples the guidance feature map $I$ to obtain a low-resolution feature map $I_l$, which has the same size of the filtering feature map $O$.
Then we minimize the reconstruction error between $I_l$ and $O$ to obtain the coefficients of the attention guided filter $A_l, B_l$,
which correspond to $I_l$.
After that, by upsampling $A_l$ and $B_l$,  the coefficients $A_h$ and $B_h$ are obtained to generate the final high-resolution output $\tilde{O}$ of the attention guided filter.
Concretely, the attention guided filter constructs a squared window $w_k$ with a radius $r$ for each position $k$. Let $I_{l_{i}}$ be a pixel of $I_l$, its output with respect to $w_k$ is obtained by a linear transformation: $\hat{O}_{ki} = a_k I_{l_i}+b_k, \forall i \in w_k$, where $a_k$ and $b_k$ are the linear coefficients of the window $w_k$.

To determine the linear coefficients $(a_k, b_k)$, we minimize the difference between $\hat{O}_{ki}$ and $O_i$ for all the pixels in the window $w_k$, which is formulated as the following optimization problem:
\begin{equation}
\min_{a_k, b_k}
E(a_k,b_k) := \sum_{i \in w_k}(T_{i}^2(a_kI_{l_i}+b_k-O_i)^2+\lambda a_k^2),
\label{equ2}
\end{equation}
where $\lambda$ is a regularization parameter, and $T_i$ is the attention weight at the position $i$.
The closed-form solution to Problem (\ref{equ2}) is given as:
\begin{equation}
a_k = \frac{\overline{T_{i}^2I_{i}O_{i}} - N_k\times\overline{X_{i}T_{i}I_{i}}\times\overline{T_{i}O_{i}}}{\overline{T_{i}^2I_{i}^2}-N_k\times\overline{X_{i}T_{i}I_{i}}\times\overline{T_{i}I_{i}}+\lambda}
, \quad b_k = \frac{\overline{T_{i}O_{i}}-a_k\times\overline{T_{i}I_{i}}}{\overline{T_{i}}},
\end{equation}
where $N_k$ is the number of the pixels in $w_k$, $X_i= \frac{T_i}{\sum_{i \in w_k}{T_i}}$, and $\overline{(\cdot)}$ is the mean of $(\cdot)$.
Considering that each position $i$ is involved in multiple windows $\{w_k\}$ with different coeffecients $\{ a_k, b_k \}$,
we average all the values of $\hat{O}_{ki}$ from different windows
to generate $\hat{O}_i$,
which is equal to average the coefficients $(a_k, b_k)$ of all the windows overlapping $i$, as following,
\begin{equation}
\hat{O}_i = \frac{1}{N_k}\sum_{k \in \Omega_i}a_kI_i+\frac{1}{N_k}\sum_{k \in \Omega_i}b_k
= A_l * I_l + B_l,
\label{eq6}
\end{equation}
where $\Omega_{i}$ is the set of all the windows including the position $i$, and $*$ is the element-wise multiplication.
After upsampling $A_l$ and $B_l$ to obtain $A_h$ and $B_h$, respectively, the final output is calculated as
$\tilde{O} = A_h * I+B_h $.

\begin{figure}[!t]
	\centering
	\includegraphics[width=0.8\linewidth]{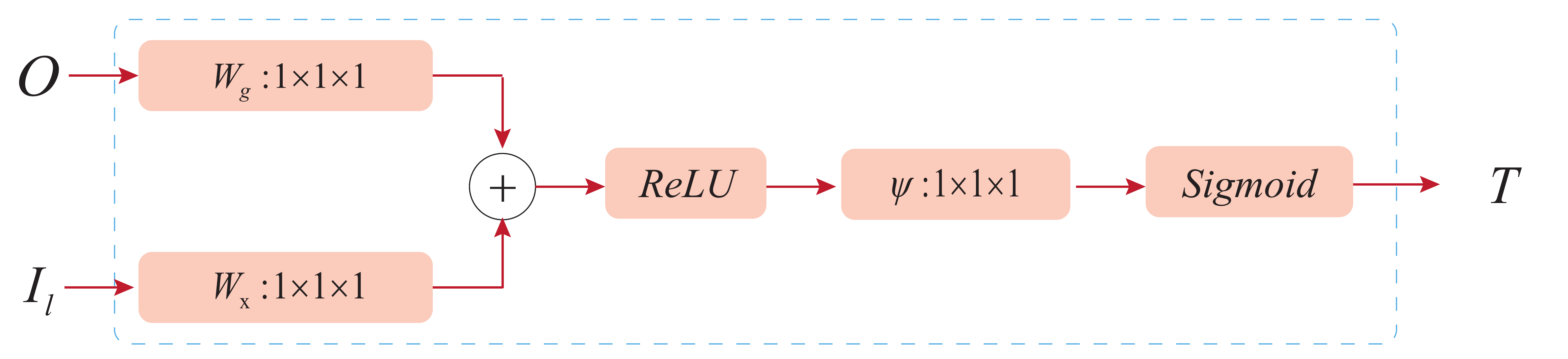}
	\caption{Schematic of the attention block. $O$ and $I$ are the inputs of attention guided filter and $T$ is the calculated attention map.}
	\label{figAttention}
\end{figure}

\subsubsection{Attention Block} is very essential in our method. Specially, the attention block is used to highlight the foreground and reduce the effect of background. As shown in Fig.~\ref{figAttention}, the attention block consists of three steps: 1) given the feature maps $O,I_l \in\mathbb{R}^{C \times H \times W}$, a channel-wise $1\times1\times1$ convolution is used to do a linear transformation.  Note that this can be referred to as the vector concatenation-based attention~\cite{wang2018non}, where the concatenated features are linearly mapped into a latent space. 2) two transformed feature maps are combined with element-wise addition with a ReLU layer. 3)  a $1\times1\times1$ convolution is applied as a additional linear transformation with a Sigmoid activation to produce the final attention map $T$.

\section{Experiments}
\label{Sec:Experiments}
In this paper, we evaluate our method in two major tasks of vessel segmentation, and optic disc/cup segmentation from retina fundus images.

\subsection{Vessel Segmentation on DRIVE Dataset}
\label{Sec:DRIVEResult}

We conduct vessel segmentation experiments on DRIVE to evaluate performance of our proposed AG-Net. The DRIVE \cite{staal2004ridge} (Digital Retinal Images for Vessel Extraction) dataset contains 40 colored fundus images,
which are obtained from a diabetic retinopathy screening program in Netherlands. The 40 images are divided into 20 training images and 20 testing images.
All the images are made by a 3CCD camera and each has size of $565\times584$. We apply gamma correction to improve the image quality, and resize the preprocessed images into $512 \times 512$ as inputs. In the experiment, we train our AG-Net from scratch using Adam with the learning rate of 0.0015. The batch size is set to 2.  The radius of windows $r$ and the regularization parameter $\lambda$ in attention guided filter are set to $2$ and $0.01$ respectively. Following the previous work \cite{zhang2018deep}, we employ Specificity (Spe), Sensitivity (Sen), Accuracy (Acc), intersection-over-union(IOU) and Area Under ROC (AUC) as measurements.

We compare our AG-Net with several state-of-the-art methods, including Li~\cite{li2016cross}, Liskowski~\cite{liskowski2016segmenting}, and Zhang~\cite{zhang2018deep}. Li~\cite{li2016cross} remolded the task of segmentation as a problem of cross-modality data transformation from retinal image to vessel map, and outputted the label map of all pixels instead of a single label of the center pixel. Liskowski~\cite{liskowski2016segmenting} trained a deep neural network on sample of examples preprocessed with global contrast normalization, zero-phase whitening, and augmented using geometric transformations and gamma corrections.
MS-NFN~\cite{wu2018multiscale} generates multi-scale feature maps with an `up-pool' submodel and a `pool-up' submodel. To verify the efficacy of attention in guided filter and transfer structural information, we replaced the attention guided filter in AG-Net with the original guided filter, named  GF-Net.

\begin{table}[!t]
	\centering
	\caption{Quantitative comparison of segmentation results on DRIVE}
	\begin{tabular}{cccccc}
		\toprule
		                 Method                  &        Acc~~        &        AUC~~        &        Sen~~        &        Spe~~        &        IOU~~        \\ \midrule
		         Li \cite{li2016cross}           &      0.9527~~       &      0.9738~~       &      0.7569~~       &      0.9816~~       &        $-~~$        \\
		Liskowski \cite{liskowski2016segmenting} &      0.9535~~       &      0.9790~~       &      0.7811~~       &      0.9807~~       &        $-~~$        \\
		      MS-NFN~\cite{wu2018multiscale}       &     $0.9567~~$      &     $0.9807~~$      & $0.7844~~$ &     $0.9819~~$      &        $-~~$        \\
		                 U-Net~\cite{ronneberger2015u}                   &     $0.9681~~$      &     $0.9836~~$      &     $0.7897~~$      &     $0.9854~~$      &     $0.6834~~$      \\
		                 M-Net~\cite{fu2018joint}                   &     $0.9674~~$      &     $0.9829~~$      &     $0.7680~~$      & $\mathbf{0.9868}~~$ &     $0.6726~~$      \\ \midrule
		            GF-Net           &     $0.9682~~$      &     $0.9837~~$      &     $0.7895~~$      &     $0.9856~~$      &     $0.6839~~$      \\
		             AG-Net                & $\mathbf{0.9692}~~$ & $\mathbf{0.9856}~~$ &     $\mathbf{0.8100}~~$      &     $0.9848~~$      & $\mathbf{0.6965}~~$ \\ \bottomrule
	\end{tabular}
	\label{DRIVE}
\end{table}

Table~\ref{DRIVE} shows the performances of different methods on DRIVE. Form the results, we could have several interesting observations: Firstly, GF-Net performs better than original M-Net, which demonstrates the superiority of the guided filter compared to the skip connection for transferring structural information.
Secondly, AG-Net outperforms GF-Net by 0.0010, 0.0019, 0.0205 and 0.0126 in terms of Acc, AUC, Sen and IOU respectively. This demonstrates the effectiveness of the attention strategy in attention guided filter.
Lastly, unlike other deep learning methods which crop images into patches, our method achieves the best performance with the original preprocessed 20 images.
We draw similar observations from the results on the CHASE\_DB1 dataset, which are shown in Table \ref{CHASE}.



\begin{figure}[!t]
	\centering
	\includegraphics[width=1\linewidth]{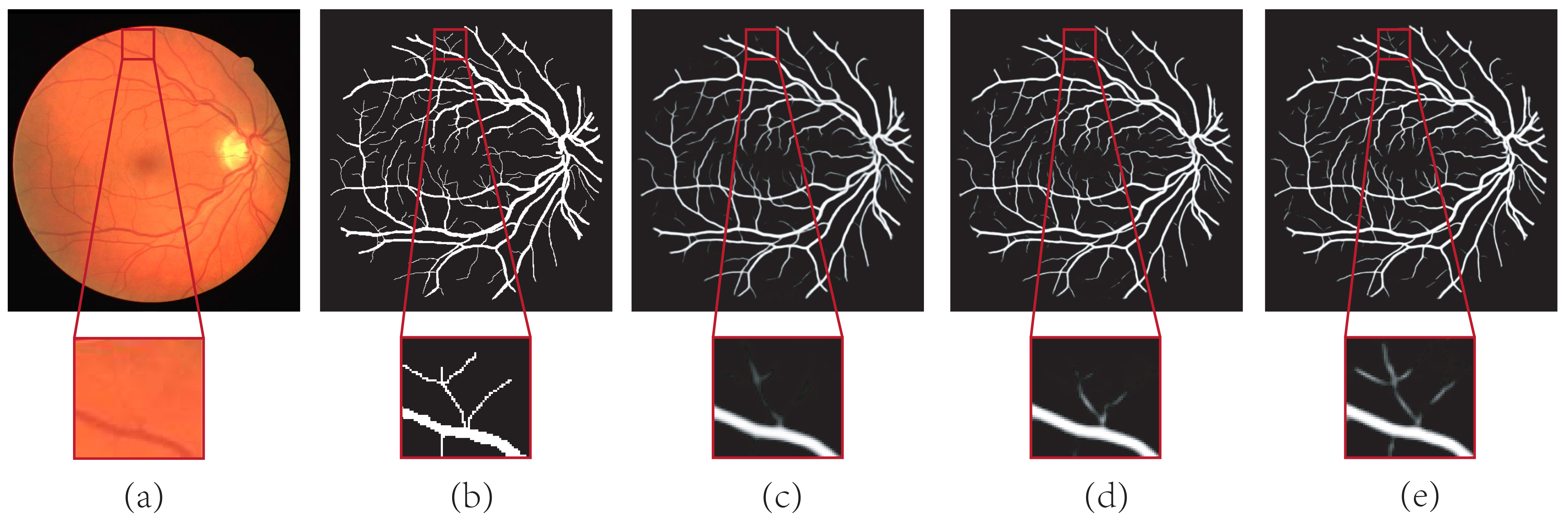}
	\caption{(a) A test image from DRIVE dataset; (b) Ground truth segmentation; (c) Segmentation result by M-Net; (d) Segmentation result by GF-Net; (e) Segmentation result by AG-Net. From (c), M-Net neglect some edge structures which are very similar to choroidal vessels. On the contrary, by exploiting attention guided as a special expanding path, AG-Net gains better discrimination power and is able to distinguish objects from similar structures. Moreover, GF helps to obtain clearer boundaries.}
	\label{figresults}
\end{figure}

Fig. \ref{figresults} shows an example test,
including the ground truth vessel and the segmentation results obtained by M-Net, M-Net+GF and the proposed AG-Net.
M-Net+GF produces clearer boundaries than M-Net,
which demonstrates the effectiveness of the guided filter to better leverage structure information.
Compared with M-Net+GF,
our proposed AG-Net produces
more precise segmentation boundaries,
which verifies that the attention mechanism is able to highlight the foreground and reduce the effect of background.


\begin{table}[!t]
	\centering
	\caption{Quantitative comparison of segmentation results on CHASE\_DB1}
	\begin{tabular}{cccccc}
		\toprule
		                 Method                  &        Acc~~        &        AUC~~        &        Sen~~        &        Spe~~        &        IOU~~        \\ \midrule
		         Li \cite{li2016cross}           &      0.9581~~       &      0.9716~~       &      0.7507~~       &      0.9793~~       &        $-~~$        \\
		Liskowski \cite{liskowski2016segmenting} &      0.9628~~       &      0.9823~~       &      0.7816~~       &      0.9836~~       &        $-~~$        \\
		      MS-NFN~\cite{wu2018multiscale}       &     $0.9637~~$      &     $0.9825~~$      & $0.7538~~$ &     $0.9847~~$      &        $-~~$        \\
		                 U-Net~\cite{ronneberger2015u}                   &     $0.9723~~$      &     $0.9837~~$      &     $0.7715~~$      &     $\mathbf{0.9858}~~$      &     $0.6366~~$      \\
		                 M-Net~\cite{fu2018joint}                   &     $0.9729~~$      &     $0.9845~~$      &     $0.7922~~$      & $0.9851~~$ &     $0.6483~~$      \\ \midrule
		            GF-Net           &     $0.9734~~$      &     $0.9853~~$      &     $0.8089~~$      &     $0.9845~~$      &     $0.6572~~$      \\
		             AG-Net                & $\mathbf{0.9743}~~$ & $\mathbf{0.9863}~~$ &     $\mathbf{0.8186}~~$      &     $0.9848~~$      & $\mathbf{0.6669}~~$ \\ \bottomrule
	\end{tabular}
	\label{CHASE}
\end{table}

In terms of time consumption, we compare our AG-Net with M-Net which is the backbone of our method. In our experiment, both algorithms are implemented with Pytorch and tested on a single NVIDIA Titan X GPU (200 iterations on DRIVE dataset). The running time is shown in Table \ref{RuningTime}.

\begin{table}[!t]
	\centering
	\caption{Quantitative comparison of the time consumption}

	\begin{tabular}{lcc}
		\toprule
		Method~~~ & $Train~time~(s)~~~$ & $Test~time~(s/image)~~~$ \\
		\midrule
		M-Net~~~ & $1800~~~$ & $0.0691~~~$ \\
		AG-Net~~~ & $2800~~~$ & $0.0158~~~$ \\
		\bottomrule
	\end{tabular}
	\label{RuningTime}
\end{table}

\subsection{Optic Dice/Cup Segmentation on ORIGA Dataset}
\label{Sec:ORIGAResult}

Optic Dice/Cup Segmentation is another important retinal segmentation task. In this experiment, we use  ORIGA dataset, which contains 650 fundus images with 168 glaucomatous eyes and 482 normal eyes. The 650 images are divided into 325 training images (including 73 glaucoma cases) and 325 testing images (including 95 glaucoma cases). We crop the OD area and resize it into $256\times256$ as the input. The training setting of our AG-Net is as same as in vessel segmentation task. We compare AG-MNet with several state-of-the-art methods in OD and/or OC segmentation, including ASM \cite{yin2011model}, Superpixel \cite{cheng2013superpixel}, LRR \cite{xu2014optic}, U-Net \cite{ronneberger2015u}, M-Net \cite{fu2018joint}, and M-Net with polar transformation (M-Net + PT).
ASM \cite{yin2011model} employs the circular hough transform initializaiton to segmentation. Superpixel method in \cite{cheng2013superpixel} utilizes superpixel classification to detect the OD and OC boundaries. The methods in LRR \cite{xu2014optic} obtain good results, but it only focus on OC segmentation.

Following the setting in~\cite{fu2018joint}, we firstly localize the disc center, and then crop $640\times640$ pixels to obtain the input images. Inspired by M-Net+PT~\cite{fu2018joint}, we provide the results of AG-Net with polar transformation, called AG-MNet+PT. Besides, to reduce the impacts of changes in the size of OD, we construct a method AG-MNet+PT$^*$, which enlarges 50 pixels of bounding-boxes in up, down, right and left, where the bounding boxes are obtained from pretrained LinkNet\cite{chaurasia2017linknet}.
We employ overlapping error (OE) as the evaluation metric, which is defined as $OE=1-\frac{A_{GT}\bigcap A_{SR}}{A_{GT}\bigcup A_{SR}}$, where $A_{GT}$ and $A_{SR}$ denote ground truth area and segmented mask, respectively.
In particular, $OE_{disc}$ and $OE_{cup}$ are the overlapping error of OD and OE. $OE_{total}$ is the average of $OE_{disc}$ and $OE_{cup}$.

\begin{table}[!t]
	\centering
	\caption{Quantitative comparison of segmentation results on ORIGA}

	\begin{tabular}{llllr}
		\toprule
		Method & $OE_{disc}~~~$ & $OE_{cup}~~~$ & $OE_{total}$\\
		\midrule
		ASM \cite{yin2011model} & $0.148$ & $0.313$ & $0.231$\\
		SP \cite{cheng2013superpixel} & $0.102$ & $0.264$ & $0.183$\\
		LRR \cite{xu2014optic} & $-$ & $0.244$ & $-$\\
		U-Net \cite{ronneberger2015u} & $0.115$ & $0.287$ & $0.201$\\
		M-Net \cite{fu2018joint} & $0.083$ & $0.256$ & $0.170$\\
		M-Net+PT \cite{fu2018joint} & $0.071$ & $0.230$ & $0.150$\\
		AG-Net (ours) & $0.069$ & $0.227$ & $0.148$\\
		AG-Net+PT (ours) & $0.067$ & $0.217$ & $0.142$\\
		AG-Net+PT$^*$ (ours) & $\mathbf{0.061}$ & $\mathbf{0.212}$ & $\mathbf{0.137}$ \\
		\bottomrule
	\end{tabular}
	\label{Table-ORIGA}
\end{table}

Table \ref{Table-ORIGA} shows the segmentation results,
where the overlapping errors of other approaches come directly from the published results. Our method outperforms all the state-of-the-art OD and/or OC segmentation algorithms in terms of the aforementioned two evaluation criteria, which demonstrates the effectiveness of our model. Besides, Our AG-Mnet performs much better than original M-Net under the same situation, which further demonstrates our attention guided filter is beneficial for the segmentation performance.
More visualization results could be found in Supplementary Material.


\section{Conclusions}

In this paper, we propose an attention guided filter as a structure sensitive expanding path. Specially, we employ M-Net as the main body and exploit our attention guided filter to replace the skip-connection and upsampling, which brings better information fusion. In addition, by introducing the attention mechanism into the guided filter, the attention guided filter can highlight the foreground and reduce the effect of background.
Experiments on two tasks demonstrate the effectiveness of our method.

\small{
\noindent
\textbf{Acknowledments.} This work was supported by National Natural Science Foundation of China (NSFC) 61602185 and 61876208, Guangdong Introducing Innovative and Enterpreneurial Teams 2017ZT07X183, and Guangdong Provincial Scientific and Technological Fund 2018B010107001, 2017B090901008 and 2018B010108002, and Pearl River S\&T Nova Program of Guangzhou 201806010081, and CCF-Tencent Open Research Fund RAGR20190103.}

\bibliographystyle{IEEEtran}
\footnotesize{\bibliography{paper992}} 
\end{document}